\documentclass[letterpaper, 10 pt, conference]{ieeeconf}  

\IEEEoverridecommandlockouts                              

\let\labelindent\relax	

\usepackage{amsmath,amssymb,amsfonts}
\usepackage{mathabx}
\usepackage{algorithmic}
\usepackage{graphicx}
\usepackage{textcomp}
\usepackage[utf8]{inputenc}
\usepackage[T1]{fontenc}
\usepackage{amsmath}
\usepackage{amsfonts}
\usepackage{amssymb}
\usepackage{balance}
\usepackage{graphicx}
\usepackage[rightcaption]{sidecap}
\usepackage{import}
\usepackage{textcomp}
\usepackage[dvipsnames]{xcolor}
\usepackage{bbm}
\usepackage{slashed}
\usepackage{caption}
\usepackage{subcaption}
\usepackage{lipsum}
\usepackage{enumitem}
\usepackage{tabularx}
\usepackage[noadjust]{cite}
\usepackage{easyReview}
\usepackage{bm}
\usepackage{tikz}
\usepackage{textcomp}

\newcommand\copyrighttext{%
	\footnotesize \textcopyright 2024 IEEE. Personal use of this material is permitted.
	Permission from IEEE must be obtained for all other uses, in any current or future
	media, including reprinting/republishing this material for advertising or promotional
	purposes, creating new collective works, for resale or redistribution to servers or
	lists, or reuse of any copyrighted component of this work in other works.
	}
\newcommand\copyrightnotice{%
	\begin{tikzpicture}[remember picture,overlay]
		\node[anchor=south,yshift=10pt] at (current page.south) {\fbox{\parbox{\dimexpr\textwidth-\fboxsep-\fboxrule\relax}{\copyrighttext}}};
	\end{tikzpicture}%
}

\newcommand{\bbR}{\mathbb{R}}

\newcommand{\calC}{\mathcal{C}}
\newcommand{\calD}{\mathcal{D}}

\newcommand{\calG}{\mathcal{G}}
\newcommand{\calH}{\mathcal{H}}

\newcommand{\calK}{\mathcal{K}}

\newcommand{\calU}{\mathcal{U}}

\newcommand{\calX}{\mathcal{X}}

\usepackage{amsthm}

\theoremstyle{definition}
\newtheorem{theorem}{Theorem}
\newtheorem{lemma}[theorem]{Lemma}
\newtheorem{proposition}[theorem]{Proposition}
\newtheorem{corollary}[theorem]{Corollary} 
\newtheorem{definition}{Definition}

\theoremstyle{remark}
\newtheorem{remark}{Remark}

\title{\LARGE \bf
From Time-Invariant to Uniformly Time-Varying Control Barrier Functions: A Constructive Approach
}

\author{Adrian Wiltz and Dimos V. Dimarogonas
\thanks{This work was supported by the ERC Consolidator Grant LEAFHOUND, the Horizon Europe EIC project SymAware (101070802), the Swedish Research Council, and the Knut and Alice Wallenberg Foundation.}
\thanks{The authors are with the Division of Decision and Control Systems, KTH Royal Institute of Technology, SE-100 44 Stockholm, Sweden
	{\tt\small \{wiltz,dimos\}@kth.se}.}%
}

\begin{document}

\maketitle
\copyrightnotice
\thispagestyle{empty}
\pagestyle{empty}


\begin{abstract}
In this paper, we define and analyze a subclass of (time-invariant) Control Barrier Functions (CBF) that have favorable properties for the construction of uniformly time-varying CBFs and thereby for the satisfaction of uniformly time-varying constraints. We call them $ \Lambda $-shiftable CBFs where $ \Lambda $ states the extent by which the CBF can be varied by adding a time-varying function. Moreover, we derive sufficient conditions under which a time-varying CBF can be obtained from a time-invariant one, and we propose a systematic construction method. Advantageous about our approach is that a $ \Lambda $-shiftable CBF, once constructed, can be reused for various control objectives. In the end, we relate the class of $ \Lambda $-shiftable CBFs to Control Lyapunov Functions (CLF), and we illustrate the application of our results with a relevant simulation example.
\end{abstract}


\section{Introduction}

\emph{Control Barrier Functions} (CBF) as introduced in~\cite{Ames2017} are a well-established and useful tool for ensuring the forward invariance of sets for dynamical systems. They allow for constraint satisfaction and the construction of safety filters~\cite{Wabersich2023a,Hobbs2023}. Many works on (zeroing) CBFs focus on zero super-level sets, which are the set of states where a CBF takes non-negative values. As zero super-level sets often represent constraints, they are also referred to as \emph{safe set}. However, also on the so often neglected neighborhood of the zero super-level sets, where the CBF takes negative values, the CBF possesses appealing properties: it ensures the asymptotic convergence of states outside of the safe set back into it~\cite{Ames2017}; and it gives rise to the construction of time-varying CBFs from a time-invariant CBF which is the topic of this paper. 

Due to the close link between the construction of CBFs and reachability analysis~\cite{Choi2021,Squires2018,Wiltz2023a}, the construction of CBFs can be computationally expensive. Especially the construction of CBFs for time-varying constraints is problematic as this adds another dimension to the state space and the CBF needs to be computed on a possibly infinitely long time-interval in order to guarantee constraint satisfaction~\cite{Choi2021}. This often does not result in a tractable problem. We aim here at mitigating this problem by constructing a time-varying CBF for certain classes of uniformly time-varying constraints on the basis of a time-invariant CBF. 

The problem of ensuring the forward invariance of time-varying sets occurs in many contexts: continuously time-varying constraints may naturally arise from the problem formulation as transient performance specifications or time-varying physical limitations, but they may also occur as time-invariant constraints that only need to be eventually satisfied after some time. A rather general class of time-varying constraints are spatio-temporal logic constraints, for example specified in terms of the STL (Signal Temporal Logic) formalism~\cite{Maler2004}, which are state and time constraints that are combined via logic statements with each other. In~\cite{Lindemann2019,Wiltz2022a,Garg2019a}, approaches to ensure the satisfaction of broad classes of STL constraints have been proposed. While the latter work is based on (fixed time) Control Lyapunov Functions (CLF), the first two works are based on time-varying CBFs. Indeed, CBFs and CLFs appear to be suitable tools for ensuring such types of time-varying constraints as they characterize the (local) controllability properties of dynamical systems. 

In this paper, we construct and analyze a class of time-varying CBFs for input constrained systems that allow for the satisfaction of uniformly time-varying constraints. To this end, we first characterize those time-invariant CBFs~$ b(x) $ defined with respect to some given input constrained system, for which there exists a trajectory $ \bm{\lambda}(t) $ such that 
\begin{align*}
	B_{\bm{\lambda}(\cdot)}(t,x) := b(x) + \bm{\lambda}(t)
\end{align*}
constitutes a time-varying CBF with respect to the same input constrained system. Based on this, we provide sufficient conditions for trajectories $ \bm{\lambda} $ such that $ B_{\bm{\lambda}(\cdot)}(t,x) $ is guaranteed to be a CBF.
The presented results give rise to a systematic construction of CBFs of the form $ b(x) + \bm{\lambda}(t) $. Such CBFs allow for the satisfaction of uniformly time-varying constraints such as
\begin{align}
	\label{eq:time-varying constraint}
	x(t)\in\calH(t) := \{x\, | \, h(x) \geq -\bm{\lambda}(t) \} \qquad \forall t\geq0
\end{align}
where $ h: \bbR^{n} \rightarrow \bbR $ is some Lipschitz-continuous function, but also of constraints with more general uniformly time-varying sets $ \widetilde{\calH}(t) := \{ x\, | \, \tilde{h}(t,x)\geq 0\} $ where $ \widetilde{\calH}(t)\subseteq\calH(t) $. 

Constraints of the form~\eqref{eq:time-varying constraint} are essential to approaches that encode STL specifications into CBFs~\cite{Lindemann2019,Wiltz2022a}. These works however only establish under some controllability condition that $ B_{\bm{\lambda}(\cdot)} $ is a CBF which is not constructive. With our result, we resolve this problem by providing a constructive approach to the synthesis of $ B_{\bm{\lambda}(\cdot)} $ as a CBF. Furthermore, our method allows for the satisfaction of uniformly time-varying constraints even for systems that cannot be brought to Byrnes-Isidori form~\cite{Byrnes1984} which is in contrast to the high gain approach presented in~\cite{Mehdifar2023}. Our results hold irrespectively of the particular form of the underlying dynamics. 

The remainder is structured as follows. Section~\ref{sec:preliminaries} introduces preliminaries; Section~\ref{sec:main results} introduces a subclass of CBFs that has favorable properties for the construction of uniformly time-varying CBFs, and analyzes it in detail; Section~\ref{sec:relation of lambda shiftable cbf and clf} relates this class of CBFs to CLFs; Section~\ref{sec:simulation} presents some simulation examples, and a conclusion is drawn in Section~\ref{sec:conclusion}.

\emph{Notation:} A continuous, strictly increasing function $ \alpha: \bbR_{\geq0} \rightarrow \bbR_{\geq0} $ with $ \alpha(0) = 0 $ is called a class $ \calK $ function, and if the function is additionally defined on the entire $ \bbR $ such that $ \alpha: \bbR \rightarrow \bbR $ then it is called an extended class $ \calK_{e} $ function. A trajectory $ \bm{x}: \bbR \rightarrow \calX $ is denoted with boldface, and $ \bm{\calX}_{[t_{1},t_{2}]} $ denotes the set of all such trajectories defined on $ [t_{1},t_{2}]\subseteq \bbR $.

\section{Preliminaries}
\label{sec:preliminaries}

Throughout this paper, we consider the input-constrained dynamical system
\begin{align}
	\label{eq:dynamics}
	\dot{x} = f(x,u), \qquad x(0) = x_{0},
\end{align}
where $ x, x_{0}\in\bbR^{n} $, $ u\in\calU\subseteq\bbR^{m} $, and $ f: \bbR^{n}\times\calU \rightarrow\bbR^{n} $ is Lipschitz-continuous in both of its arguments in order to ensure the uniqueness of its arguments; forward completeness is assumed. The solution to~\eqref{eq:dynamics} for an input trajectory $ \bm{u}: \bbR_{\geq0} \rightarrow \bm{\calU} $ is denoted by $ \bm{\varphi}(t;x_{0},\bm{u}) $; the first argument~$ t $ denotes the time at which $ \bm{\varphi} $ is evaluated. Let us now formally define CBFs.

\begin{definition}(CBF, similar to~\cite[Def.~5]{Ames2017})
	\label{def:cbf}
	Consider $ \calD\subseteq\bbR^{n} $ and a continuously differentiable function $ b: \bbR^{n}\rightarrow\bbR $ such that $ \calC\subseteq\calD\subseteq\bbR^{n} $, where 
	\begin{align}
		\label{eq:calC}
		\calC := \{x \, | \, b(x)\geq 0\}
	\end{align}
	is the zero super-level set of $ b $. We call $ b $ a \emph{Control Barrier Function} (CBF) on $ \calD $ with respect to~\eqref{eq:dynamics} if there exists an extended class $ \calK_{e} $ function $ \alpha $ such that for all $ x\in\calD $ 
	\begin{align}
		\label{eq:def cbf 1}
		\sup_{u\in\calU} \left\{ \frac{\partial b}{\partial x}(x) \, f(x,u) \right\} \geq -\alpha (b(x)). 
	\end{align}
\end{definition}

Next, let us consider the (time-invariant) state constraint 
\begin{align}
	\label{eq:time-invarinat state constraint}
	x(t) \in \calH := \{x \, | \, h(x)\geq 0\} \qquad \forall t\geq 0
\end{align}
where $ h: \bbR^{n}\rightarrow\bbR $ is a Lipschitz-continuous function. A CBF $ b $ can be viewed as a system theoretic characterization of a dynamical system~\eqref{eq:dynamics} with respect to a constraint~\eqref{eq:time-invarinat state constraint} when it is chosen as $ b(x) \leq h(x) $ for all $ x\in\calC $. Then, $ \calC\subseteq\calH $ holds for the zero super-level set of the CBF.

We call the set $ \calC $ \emph{forward control invariant} with respect to system~\eqref{eq:dynamics} if there exist $ \bm{u}\in\bm{\calU}_{[0,\infty)} $ such that $\bm{\varphi}(t;x_{0},\bm{u})\in\calC$ for all $ t\geq 0 $. Moreover, we call the set $ \calC $ \emph{forward invariant} under input $ \bm{u}\in\bm{\calU}_{[0,\infty)} $ with respect to system~\eqref{eq:dynamics} if $\bm{\varphi}(t;x_{0},\bm{u})\in\calC$ for all $ t\geq 0 $. 
The following forward invariance result for CBFs is a corollary of Nagumo's theorem~\cite{Nagumo1942}.

\begin{corollary}[Forward invariance via CBFs]
	\label{corollary:cbf invariance}
	Let $ b $ be a CBF to~\eqref{def:cbf} on $ \calD\subseteq\bbR^{n} $. Then, any locally Lipschitz continuous control $ u(x)\in\calK_{\text{CBF}}(x):=\{ u\in\calU \, | \, \frac{\partial b}{\partial x}(x) \, f(x,u) \geq -\alpha (b(x)) \} $ renders $ \calC\subseteq\calD $ forward invariant.
\end{corollary}

\section{Main Results}
\label{sec:main results}

Let us introduce a more particular notion of CBFs that have favorable properties for ensuring the satisfaction of uniformly time-varying constraints.

\subsection{$ \Lambda $-shiftable CBFs}
We define this class, which constitutes a subclass of CBFs according to Definition~\ref{def:cbf}, as follows.
\begin{definition}($ \Lambda $-shiftable CBF)
	\label{def:lambda shiftable CBF}
	A continuously differentiable function $ b:\bbR^{n}\rightarrow\bbR_{\geq 0} $ is called a \emph{$ \Lambda $-shiftable CBF} with respect to~\eqref{eq:dynamics} for some $ \Lambda > 0 $ if $ b(x) $ is a CBF on the domain 
	\begin{align*}
		\calC_{\Lambda} := \{ x \, | \, b(x) \geq -\Lambda \}
	\end{align*}
	with respect to~\eqref{eq:dynamics}, or equivalently, if there exists an extended class~$ \calK_{e} $ function~$ \alpha $ such that~\eqref{eq:def cbf 1} holds for all $ x\in\calC_{\Lambda} $.
\end{definition}

Here, $ \calC_{\Lambda} $ takes the role of $ \calD $ in Definition~\ref{def:cbf} as domain. We call such a CBF $ \Lambda $-shiftable since $ b(x)\!+\!\lambda $, which is the by $ \lambda $ shifted version of $ b(x) $, is still a CBF for any $ \lambda\!\in\![0,\Lambda] $. 
\begin{proposition}
	\label{prop:lambda_constant_shifted}
	For $ \lambda\in[0,\Lambda] $, the by $ \lambda $ shifted version of $ b(x) $ defined as
	\begin{align}
		\label{eq:prop:B_lambda}
		B_{\lambda}(x) := b(x) + \lambda
	\end{align}
	is a CBF on $ \calC_{\Lambda} $. If additionally $ \lambda < \Lambda $, then $ B_{\lambda} $ is also a $ (\Lambda\!-\!\lambda) $-shiftable CBF. 
\end{proposition}
\begin{proof}
	For the first part of the proposition, we observe that 
	\begin{align*}
		&\sup_{u\in\calU} \left\{ \frac{\partial B_{\lambda}}{\partial x} (x) f(x,u) \right\} = \sup_{u\in\calU} \left\{ \frac{\partial b}{\partial x} (x) f(x,u) \right\} \\
		&\qquad \stackrel{\text{\eqref{eq:def cbf 1}}}{\geq} -\alpha(b(x)) \geq -\alpha(b(x)+\lambda) = -\alpha(B_{\lambda}(x)) 
	\end{align*}
	for all $ x\in\calC_{\Lambda} $. Thus by Definition~\ref{def:cbf}, $ B_{\lambda} $ is a CBF on $ \calC_{\Lambda} $. Since $ \calC_{\Lambda} = \{ x \, | \, b(x) \geq -\Lambda \} = \{ x \, | \, B_{\lambda}(x) \geq -(\Lambda-\lambda) \} $, $ B_{\lambda} $ is a $ (\Lambda\!-\!\lambda) $-shiftable CBF if $ \Lambda - \lambda > 0 $, or equivalently,  $ \lambda < \Lambda$, and the second part is also shown.
\end{proof}

Clearly, every $ \Lambda $-shiftable CBF is a CBF, but not every CBF is $ \Lambda $-shiftable. 
The following is an obvious consequence of Corollary~\ref{corollary:cbf invariance} and the definition of $ \Lambda $-shiftable CBFs (Definition~\ref{def:lambda shiftable CBF}).

\begin{corollary}
	Let $ b(x) $ be a $ \Lambda $-shiftable CBF. Then for an arbitrary $ \lambda\in[0,\Lambda] $, any locally Lipschitz-continuous control $ u(x)\in\calK_{\text{CBF}}^{\lambda}(x):=\{ u\in\calU \, | \, \frac{\partial b}{\partial t}(x) \, f(x,u) \geq -\alpha (b(x)+\lambda) \} $ renders $ \calC_{\lambda}:=\{ x \, | \, b(x) \geq -\lambda \} $ forward-invariant. 
\end{corollary}

\subsection{From $ \Lambda $-shiftable CBFs to uniformly time-varying CBFs}
A $ \Lambda $-shiftable CBF $ b(x) $ even exhibits favorable properties when we add a differentiable time-dependent function $ \bm{\lambda}: \bbR_{\geq 0} \rightarrow [0,\Lambda] $ instead of a constant $ \lambda $ as in~\eqref{eq:prop:B_lambda}. Let us denote the resulting time-varying function by
\begin{align}
	\label{eq:B_lambda_time-varying}
	B_{\bm{\lambda}(\cdot)}(t,x) := b(x) + \bm{\lambda}(t).
\end{align}
The CBFs constructed in \cite{Lindemann2019,Wiltz2022a} take this form. Given that $ b(x) $ is a $ \Lambda $-shiftable CBF, our objective is to design the trajectory $ \bm{\lambda}(t) $ such that $ B_{\bm{\lambda}(\cdot)}(t,x) $ implicitly becomes a CBF with respect to dynamics~\eqref{eq:dynamics} augmented with time. This ultimately allows us to ensure the forward invariance of the uniformly time-varying set
\begin{align}
	\label{eq:C_lambda}
	\calC_{\bm{\lambda}(\cdot)}(t):=\{ x \, | \, b(x) \geq -\bm{\lambda}(t) \}.
\end{align}

To this end, we impose on $ \bm{\lambda}(t) $ the condition that for all $ t\geq 0 $ it holds
\begin{align}
	\label{eq:lambda_dot condition}
	\frac{\partial \bm{\lambda}}{\partial t}(t) \geq -\alpha_{\lambda}(\bm{\lambda}(t))
\end{align}
where $ \alpha_{\lambda} $ is a class $ \calK $ function. The differential inequality~\eqref{eq:lambda_dot condition} can be interpreted as a uniform lower-bound on the derivative of~$ \bm{\lambda} $. Intuitively, it implies that the derivative of $ \bm{\lambda} $ tends to zero if the value of $ \bm{\lambda} $ tends towards the lower bound of its range which is zero. Based on this condition, we can now show that $ B_{\bm{\lambda}(\cdot)}(t,x) $ is a CBF with respect to dynamics~\eqref{eq:dynamics} augmented with time.

\begin{theorem}
	\label{thm:time-varying CBF without input constraints}
	Let $ B_{\bm{\lambda}(\cdot)}(t,x) := b(x) + \bm{\lambda}(t) $ where $ b(x) $ is a $ \Lambda $-shiftable CBF with respect to~\eqref{eq:dynamics} with input constraint $ u\in\calU\subseteq\bbR^{m} $. Moreover, let $ \bm{\lambda}: \bbR_{\geq0}\rightarrow[0,\Lambda] $ be a continuously differentiable function that satisfies~\eqref{eq:lambda_dot condition} where $ \alpha_{\lambda} $ is an either convex or concave class $ \calK $ function and it holds $ \alpha(-\xi)\leq -\alpha_{\lambda}(\xi) $ for all $ \xi\in[0,\Lambda] $. Then, $ B_{\bm{\lambda}(\cdot)}(t,x) $ is a CBF on $ \bbR_{\geq 0} \times \calC_{\Lambda} $ with respect to dynamics~\eqref{eq:dynamics} augmented by time
	\begin{align*}
		\begin{bmatrix}
			\dot{t} \\ \dot{x}
		\end{bmatrix}
		=
		\begin{bmatrix}
			1 \\ f(x,u)
		\end{bmatrix}.
	\end{align*}
	Moreover, then there exists an extended class~$ \calK_{e} $ function $ \beta $ such that for all $ (t,x)\in\bbR_{\geq 0} \times \calC_{\Lambda} $
	\begin{align*}
		\sup_{u\in\calU} \left\{ \frac{\partial B_{\bm{\lambda}(\cdot)}}{\partial (t,x)}(t,x) \, \begin{bmatrix} 1 \\ f(x,u) \end{bmatrix} \right\} \geq -\beta(b(x)+\bm{\lambda}(t)),
	\end{align*}
	where $ \frac{\partial B_{\bm{\lambda}(\cdot)}}{\partial (t,x)} := \begin{bmatrix}
		\frac{\partial B_{\bm{\lambda}(\cdot)}}{\partial t} & \frac{\partial B_{\bm{\lambda}(\cdot)}}{\partial x}
	\end{bmatrix} $.
\end{theorem}

\begin{remark}
	The condition that $ \alpha(-\xi)\leq -\alpha_{\lambda}(\xi) $ must hold for all $ \xi\in[0,\Lambda] $ has an intuitive interpretation. To this end, note that $ \alpha $ (on the left-hand side) characterizes the minimal possible ascend of the system state on the $ \Lambda $-shiftable CBF~$ b $, while $ \alpha_{\lambda} $ is bounding the ``speed'' with which~$ b $ is shifted. Consequently, the condition $ \alpha(-\xi)\leq -\alpha_{\lambda}(\xi) $ ensures that the function $ b $ is not shifted faster than the system state can move towards the interior of $ \calC_{\bm{\lambda}(\cdot)} $. 
\end{remark}

In order to formally prove Theorem~\ref{thm:time-varying CBF without input constraints}, we first need to derive some intermediate results. In particular, it is crucial to determine under which conditions there exists an extended class~$ \calK_{e} $ function~$ \beta $ that upper-bounds the sum of~$ \alpha $ and~$ \alpha_{\lambda} $. More precisely, we first need to provide sufficient conditions that guarantee the existence of an extended class~$ \calK_{e} $ function~$ \beta $ such that
\begin{align}
	\label{eq:class K function sum is upper-bounded}
	\alpha(x_{1}) + \alpha_{\lambda}(x_{2}) \leq \beta(x_{1}+x_{2}).
\end{align}
Clearly, this holds with equality if $ \alpha $ and $ \alpha_{\lambda} $ are linear. However, when such strong assumptions do not hold, more sophisticated conditions are required. These are stated in the subsequent lemmas.

\begin{lemma}[\cite{Wiltz2024}]
	\label{lemma:time-varying CBF without input constraints 1}
	Let $ \alpha_{1}: \bbR \rightarrow \bbR $ be an extended class $ \calK_{e} $ function, and $ \alpha_{2}: \bbR_{\geq 0} \rightarrow \bbR $ a \emph{convex} class $ \calK $ function such that $ \alpha_{1}(-x)\leq -\alpha_{2}(x) $ for all $ x\in[0,A] $ and some finite $ A>0 $. Then, there exists an extended class~$ \calK_{e} $ function $ \beta $ such that for all $ x_{1} \in [-A,\infty) $, $ x_{2} \in [0,A] $ it holds
	\begin{align}
		\label{eq:time-varying CBF without input constraints 1}
		\alpha_{1}(x_{1}) + \alpha_{2}(x_{2}) \leq \beta(x_{1}+x_{2}).
	\end{align}
\end{lemma}

\begin{lemma}[\cite{Wiltz2024}]
	\label{lemma:time-varying CBF without input constraints 2}
	Let $ \alpha_{1}: \bbR \rightarrow \bbR $ be an extended class $ \calK_{e} $ function, and $ \alpha_{2}: \bbR_{\geq 0} \rightarrow \bbR $ a \emph{concave} class $ \calK $ function such that $ \alpha_{1}(-x)\leq -\alpha_{2}(x) $ for all $ x\in[0,A] $ and $ A>0 $. Then, there exists an extended class~$ \calK_{e} $ function $ \beta $ such that for all $ x_{1} \in [-A,\infty) $, $ x_{2} \in [0,A] $ it holds
	\begin{align}
		\label{eq:time-varying CBF without input constraints 2}
		\alpha_{1}(x_{1}) + \alpha_{2}(x_{2}) \leq \beta(x_{1}+x_{2}).
	\end{align}
	This even holds if $ A\rightarrow\infty $.
\end{lemma}

Now, we are ready to state the proof of Theorem~\ref{thm:time-varying CBF without input constraints}.

\begin{proof}[Proof of Theorem~\ref{thm:time-varying CBF without input constraints}]
	As $ b(x) $ and $ \bm{\lambda}(t) $ are both continuously differentiable, also $ B_{\lambda} $ is continuously differentiable. Thus, in order to show that $ B_{\bm{\lambda}(\cdot)} $ is a CBF, it remains to show that~\eqref{eq:def cbf 1} holds with respect to the dynamics~\eqref{eq:dynamics} augmented by time. To start with, we recall that for the $ \Lambda $-shiftable CBF~$ b $ it holds for all $ x\in\calC_{\Lambda} $ that
	\begin{align*}
		\sup_{u\in\calU} \left\{ \frac{\partial b}{\partial x}(x) \, f(x,u) \right\} \geq -\alpha (b(x)).
	\end{align*}
	By adding $ -\alpha_{\lambda}(\bm{\lambda}(t)) $ to both sides, where $ \alpha_{\lambda} $ is the class~$ \calK $ function specified in the statement of the theorem, and applying Lemmas~\ref{lemma:time-varying CBF without input constraints 1} and~\ref{lemma:time-varying CBF without input constraints 2}, we obtain
	\begin{align}
		\label{eq:thm time-varying CBF without input constraints aux 0}
			\sup_{u\in\calU}\! \left\{\! \frac{\partial b}{\partial x}\!(x)  f(x,u) \!\right\} \!-\!\alpha_{\lambda}(\bm{\lambda}(t)) \!&\geq\! -\alpha(b(x)) \!-\! \alpha_{\lambda}(\bm{\lambda}(t))  \nonumber\\
			&\hspace{-0.3cm}\stackrel{\text{Lem.~\ref{lemma:time-varying CBF without input constraints 1},\ref{lemma:time-varying CBF without input constraints 2}}}{\geq} -\beta(b(x)\!+\!\bm{\lambda}(t)).
	\end{align}
	By further employing~\eqref{eq:lambda_dot condition}, we have
	\begin{align}
		\label{eq:thm time-varying CBF without input constraints aux 1}
		\sup_{u\in\calU} \left\{ \frac{\partial b}{\partial x}(x) \, f(x,u) \right\} + \frac{\partial \bm{\lambda}}{\partial t}(t) \geq -\beta(b(x)+\bm{\lambda}(t))
	\end{align}
	and we can finally conclude that for all $ t\in\bbR_{\geq 0} $ and $ x\in\calC_{\Lambda} $ it holds
	\begin{align*}
		\begin{split}
			&\sup_{u\in\calU}\! \left\{\! \frac{\partial B_{\bm{\lambda}(\cdot)}}{\partial (t,x)}\!(t,x)  \begin{bmatrix} 1 \\ f(x,u) \end{bmatrix} \!\right\} = \sup_{u\in\calU}\! \left\{\! \frac{\partial b}{\partial x}\!(x) f(x,u) \!+\! \frac{\partial \bm{\lambda}}{\partial t}(t) \!\right\} \\
			&\quad= \sup_{u\in\calU}\left\{ \frac{\partial b}{\partial x}(x) \, f(x,u) \right\} + \frac{\partial \bm{\lambda}}{\partial t}(t) \stackrel{\eqref{eq:thm time-varying CBF without input constraints aux 1}}{\geq} -\beta(b(x)+\bm{\lambda}(t)).
		\end{split}
	\end{align*}
	This is the CBF condition~\eqref{eq:def cbf 1} with respect to the dynamics~\eqref{eq:dynamics} augmented by time, which concludes the proof.
\end{proof}

As a direct consequence of Corollary~\ref{corollary:cbf invariance} and Theorem~\ref{thm:time-varying CBF without input constraints}, we obtain the following forward invariance result. 

\begin{corollary}
	\label{corollary:time varying set invariance with input constraints}
	Let $ B_{\bm{\lambda}(\cdot)}(t,x) := b(x) + \bm{\lambda}(t) $ where $ b(x) $ is a $ \Lambda $-shiftable CBF with respect to~\eqref{eq:dynamics} with input constraint $ u\in\calU\subseteq\bbR^{m} $, and let the same assumptions as in Theorem~\ref{thm:time-varying CBF without input constraints} hold. Then, any locally Lipschitz-continuous control 
	\begin{align*}
		&u(t,x)\in\calK_{\text{CBF}}(t,x)\\
		&:=\!\!\left\{\! u\!\in\!\calU  \bigg| \frac{\partial B_{\!\bm{\lambda}(\cdot)}}{\partial x}\!(t,\!x) \, f(x,\!u) \!\!+\!\! \frac{\partial B_{\!\bm{\lambda}(\cdot)}}{\partial t}\!(t,\!x) \!\!\geq\!\! -\beta(B_{\!\bm{\lambda}(\cdot)}(t,\!x)) \!\right\}
	\end{align*}
	renders $ \calC_{\bm{\lambda}(\cdot)}(t) $ forward invariant with respect to the dynamics~\eqref{eq:dynamics} for all $ t\geq 0 $. 
\end{corollary}
\begin{proof}
	By Theorem~\ref{thm:time-varying CBF without input constraints}, $ B_{\bm{\lambda}(\cdot)}(t,x) $ is a CBF on $ \bbR_{\geq 0} \times \calC_{\Lambda} $ with respect to~\eqref{eq:dynamics}. As furthermore $ \calC_{\bm{\lambda}(\cdot)}(t) \subseteq \calC_{\Lambda} $ for all $ t\geq 0 $ and $ \calC_{\bm{\lambda}(\cdot)} $ is the zero super-level set of $ B_{\bm{\lambda}(\cdot)} $, the conditions of Corollary~\ref{corollary:cbf invariance} are satisfied and the result follows directly. 
\end{proof}
\begin{remark}
	If $ \bm{\lambda} $ is a continuous and piecewise continuously differentiable function, then Corollary~\ref{corollary:time varying set invariance with input constraints} still holds when we only require that $ u(t,x) \in \calK_{\text{CBF}}(t,x) $ for almost all $ t\geq 0 $.
\end{remark}

\subsection{Practical construction of time-varying trajectories $ \bm{\lambda}(\cdot) $}
\label{subsec:lambda function construction guidelines}

The capability of a $ \Lambda $-shiftable CBF to be shifted by adding a trajectory $ \bm{\lambda} $, especially with respect to the ``shifting-speed'', is characterized by the extended class~$ \calK_{e} $ function~$ \alpha $. This can be seen from the premises of Theorem~\ref{thm:time-varying CBF without input constraints} as it requires
\begin{align*}
	\frac{\partial \bm{\lambda}}{\partial t}(t) \geq -\alpha_{\lambda}(\bm{\lambda}(t)) \geq \alpha(-\bm{\lambda}(t)).
\end{align*}
This condition suggests the following construction of a uniformly time-varying CBF $ B_{\bm{\lambda}(\cdot)} $. Given a $ \Lambda $-shiftable CBF~$ b $, we propose the following steps:
\begin{enumerate}
	\item[(1)] Find a linear, convex or concave class~$ \calK $ function $ \alpha_{\lambda} $ (the latter even works for $ \Lambda\rightarrow\infty $) that satisfies $ \alpha_{\lambda}(\xi)\leq -\alpha(-\xi) $ for all $ \xi\in[0,\Lambda] $.
	\item[(2)] Choose any differentiable trajectory $ \bm{\lambda}: \bbR_{\geq 0} \rightarrow [0,\Lambda] $ such that $ \frac{\partial \bm{\lambda}}{\partial t}(t) \geq -\alpha_{\lambda}(\bm{\lambda}(t)) $ holds for all $ t\geq 0 $. Intuitively, the smaller the value of $ \bm{\lambda} $, the smaller must be its decrease-rate. 
	\item[(3)] The uniformly time-varying CBF is then, according to Theorem~\ref{thm:time-varying CBF without input constraints}, given by $ B_{\bm{\lambda}(\cdot)}(t,x) := b(x) + \bm{\lambda}(t) $.
\end{enumerate}

\begin{remark}
		There are multiple ways to proceed in step~(2). For example, $ \bm{\lambda} $ can be chosen as a piecewise linear function, such that $ \partial\bm{\lambda}/\partial t $ becomes piecewise constant and the analytic verification of the differential inequality becomes straight-forward. Alternatively, the first-order differential equation $ \dot{\bm{\lambda}}(t) = -\alpha_{\lambda}(\bm{\lambda}(t)) $ can be (numerically) solved.
\end{remark}

If it holds for the time-varying zero super-level set of $ B_{\bm{\lambda}}(t,x) $ denoted by $ \calC_{\bm{\lambda}(\cdot)}(t) $ (cf.~\eqref{eq:C_lambda}) that
\begin{align*}
	\calC_{\bm{\lambda}(\cdot)}(t) \subseteq \widetilde{\calH}(t) = \{ x \, | \, \tilde{h}(t,x)\geq 0 \} \qquad \forall t\geq 0,
\end{align*}
and if $ x_{0} \in\calC_{\bm{\lambda}(\cdot)}(0) $, then the satisfaction of the constraint $ \bm{\varphi}(t;x_{0},\bm{u})\in\widetilde{\calH}(t) $ for all $ t\geq0 $ can be ensured according to Corollary~\ref{corollary:time varying set invariance with input constraints}. Clearly, finding a suitable function $ B_{\bm{\lambda}(\cdot)}(t,x) $ for a given constraint set $ \widetilde{\calH}(t) $ is generally not straightforward. Yet, it can be simplified using the following steps:
\begin{enumerate}
	\item[(a)] At first, an affine function of the form $ h(x) + \bm{\lambda}_{h}(t) $ needs to be determined such that 
	\begin{align*}
		h(x) + \bm{\lambda}_{h}(t) \leq \tilde{h}(t,x) \qquad \forall t\in\bbR_{\geq 0}, \forall x\in\widetilde{\calH}(t).
	\end{align*}
	Then, it holds $ \calH(t) := \{x \, | \, h(x) \geq -\bm{\lambda}_{h}(t)\} \subseteq \widetilde{\calH}(t) $ for all $ t\geq 0 $. 
	\item[(b)] Determine a CBF $ b(x) $ with
	\begin{align*}
		\calC := \{ x \, | \, b(x)\geq 0\} \subseteq \{x \, | \, h(x) \geq 0\}
	\end{align*}
	on a domain $ \calD\in\bbR^{n} $ such that $ \calD \supset \{x \, | \, h(x) \geq -\max_{t\geq0} \bm{\lambda}_{h}(t)\} $; the right-hand side corresponds to $ \bigcup_{t\geq 0} \calH(t) $. By Definition~\ref{def:lambda shiftable CBF}, $ b(x) $ is a $ \Lambda $-shiftable CBF.
	\item[(c)] Determine a trajectory $ \bm{\lambda}(t) $ by following steps~(1)-(3) above such that $ \calC_{\bm{\lambda}(\cdot)}(t) \subseteq \calH(t) $ for all $ t\geq 0 $
\end{enumerate}
\begin{remark}
	The affine form in step~(a) may directly arise from the problem formulation. For example when handling STL tasks as in~\cite{Lindemann2019,Wiltz2022a}, all time-varying constraints can be directly defined in affine form.
\end{remark}
\begin{remark}
	Step~(b) requires the construction of a (time-invariant) CBF which is a classical problem in the CBF literature that currently receives much attention. However, not all works on CBF construction are likewise suitable: some works only consider the construction of CBFs on the domain $ \calD = \calC $. Here, however, we require $ \calD\supset\calC $. Works that consider the construction of a CBF on $ \calD\supset\calC $ are for example \cite{Cortez2022a,Robey2020,Chen2024}. Due to the close relation of $ \Lambda $-shiftable CBFs and Control Lyapunov Functions (CLF) \cite{Artstein1983,Sontag1989} as discussed in the next section, also works on CLF construction can be considered at this point, see e.g. \cite{Baier2018,Ravanbakhsh2019} and references therein.
\end{remark}


\subsection{Discussion}

Despite the continued advances in the computation of CBFs and CLFs, their construction stays non-trivial and can be computationally expensive. Therefore, it is important that already computed CBFs and CLFs can be used for various control objectives. Whereas CBFs are so far often used for the satisfaction of one particular (time-invariant) constraint, or CLFs for the synthesis of controllers for stabilizing one particular equilibrium point (similar for the notion of fixed-time CLFs~\cite{Garg2022}), our intention is to show how a single CBF can be used for various different control objectives. The class of $ \Lambda $-shiftable CBFs gives rise to further CBFs when shifted by a constant $ \lambda $ (cf. Proposition~\ref{prop:lambda_constant_shifted}), but also when shifted by a time-varying function $ \bm{\lambda}(t) $ (cf. Theorem~\ref{thm:time-varying CBF without input constraints}). Thereby, the satisfaction of a multitude of time-invariant and uniformly time-varying constraints can be ensured based on a single CBF (cf. Section~\ref{subsec:lambda function construction guidelines}). 

\section{Relation of $ \Lambda $-shiftable CBFs and CLFs}
\label{sec:relation of lambda shiftable cbf and clf}

A Control Lyapunov Function (CLF) is a value function that possesses similar properties as a CBF. From an intuitive point of view, a CBF $ b $ ensures that for any point $ x $ within its zero super-level set $ \calC $ ($ x\in\calC $) there exists a control input~$ \bm{u} $ such that $ b $ does not decrease faster along $ \bm{\varphi}(t;x_{0},\bm{u}) $ than with some prescribed maximum rate $ \alpha $, see~\eqref{eq:def cbf 1}. For points~$ x $ outside of the zero super-level set, i.e., $ x\in\calD\smallsetminus\calC $, the condition for CBFs is a bit stricter: for such states~$ x $, the existence of a control input~$ \bm{u} $ that ensures a certain prescribed minimum ascend rate on $ b $ along $ \bm{\varphi}(t;x_{0},\bm{u}) $ is guaranteed. This leads to asymptotically stabilizing feedback controllers as known from \cite[Prop.~2]{Ames2017}. A property similar to the latter one is also guaranteed to hold for CLFs. However for a CLF, it must hold in any state $ x $ of the domain where the CLF is defined. In particular, CLFs are defined as follows. 

\begin{definition}[Control Lyapunov Function (CLF); \cite{Sontag1998}, Def.~5.7.1]
	A twice continuously differentiable and positive definite function $ V $ is called a \emph{Control Lyapunov Function} (CLF) on $ \calD $ with respect to \eqref{eq:dynamics}, where $ \calD\subseteq\bbR^{n} $ is some neighborhood of the origin, if $ V $ is proper on $ \calD $ and there exists a class $ \calK $ function $ \gamma $ such that for all $ x\in\calD $
	\begin{align}
		\label{eq:def clf 1}
		\sup_{u\in\calU}\left\{ \frac{\partial V}{\partial x}(x) \; f(x,u) \right\} \leq -\gamma(V(x)).
	\end{align}
\end{definition}

As we note in the next theorem, CLFs can be directly transformed into $ \Lambda $-shiftable CBFs.

\begin{theorem}
	\label{thm:clf and cbf}
	Let $ V $ be a CLF on $ \calD\subseteq\bbR^{n} $ with respect to~\eqref{eq:dynamics}, and let $ \Lambda_{\text{max}} $ be the maximum value of the CLF~$ V $ on its largest closed sub-level set in the domain $ \calD $, that is $ \Lambda_{\text{max}} := \max\{\Lambda \,|\, \calC^{V}_{\Lambda}:=\{x \, | \, V(x) \leq \Lambda \} \subseteq \calD\} $. Furthermore, consider 
	\begin{align}
		\label{eq:thm:clf based cbf}
		b(x) := -V(x) + b_{c}
	\end{align}
	where $ b_{c}\in[0,\Lambda_{\text{max}}) $ is some constant. Then $ b(x) $ is a $ \Lambda $-shiftable CBF for any $ \Lambda\in(0,\Lambda_{\text{max}}-b_{c}] $ if $ \Lambda_{\text{max}} $ is finite, and otherwise for any $ \Lambda\in (0,\infty) $. Moreover, if $ \calD=\bbR^{n} $, then $ b $ is a $ \Lambda $-shiftable CBF with $ \Lambda\in(0,\infty) $. 
\end{theorem}
\begin{proof}
	According to Definition~\ref{def:lambda shiftable CBF}, the function $ b $ is a $ \Lambda $-shiftable CBF if for all $ x\in\calC_{\Lambda} $ it holds that
	\begin{align}
		\label{eq:thm:clf based cbf aux 0}
		\begin{split}
			\sup_{u\in\calU}\left\{\frac{\partial b}{\partial x}(x) \, f(x,u) \right\} \stackrel{\eqref{eq:thm:clf based cbf}}{=} \sup_{u\in\calU}\left\{-\frac{\partial V}{\partial x}(x) \, f(x,u) \right\} \\
			\geq -\alpha(-V(x)+b_{c})
		\end{split}
	\end{align} 
	for some extended class $ \calK_{e} $ function $ \alpha $. 
	Starting with the left-hand side of the latter equation, we derive
	\begin{align}
		\label{eq:thm:clf based cbf aux 1}
		\begin{split}
			\sup_{u\in\calU}\left\{-\frac{\partial V}{\partial x}(x) \, f(x,u) \right\} &\stackrel{\eqref{eq:def clf 1}}{\geq} \gamma(V(x)) \geq \alpha(V(x)) \\ 
			&\geq \alpha(V(x)-b_{c})
		\end{split} 
	\end{align} 
	where $ \alpha $ is a suitably constructed extended class $ \calK_{e} $ function which we choose as $ \alpha(x) = \left\{ \begin{smallmatrix}
		\gamma(x) & \text{if } x\geq 0, \hfill \\
		-\gamma(-x) & \text{if } x<0 \hfill
	\end{smallmatrix}\right. $. Next, we note that for this choice of $ \alpha $ it holds $ \alpha(x) = -\alpha(-x) $. Thus, we further obtain for the right-hand side of \eqref{eq:thm:clf based cbf aux 1} that $ \alpha(V(x)-b_{c}) = -\alpha(-V(x)+b_{c}) $. Substituting this into~\eqref{eq:thm:clf based cbf aux 1} yields~\eqref{eq:thm:clf based cbf aux 0}. As~\eqref{eq:def clf 1} only holds on $ \calD $, the largest closed super-level set of $ b $ where the CBF gradient condition~\eqref{eq:thm:clf based cbf aux 0} holds is then given by $ \calC_{\Lambda_{\text{max}}-b_{c}} = \{x\, | \, b(x) \geq -\Lambda_{\text{max}}+b_{c}\} $ if $ \Lambda_{\text{max}} $ is finite, and otherwise by $ \lim_{\Lambda\rightarrow\infty} \calC_{\Lambda} = \bbR^{n} $. Thus, we conclude that $ b $ is $ \Lambda $-shiftable for any $ \Lambda\in(0,\Lambda_{\text{max}}-b_{c}] $ if $ \Lambda_{\text{max}} $ is finite, and otherwise for any $ \Lambda \in (0,\infty) $. 
\end{proof}

\section{Simulation}
\label{sec:simulation}

\begin{figure}[t]
	\centering
	\def\svgwidth{0.85\columnwidth}
	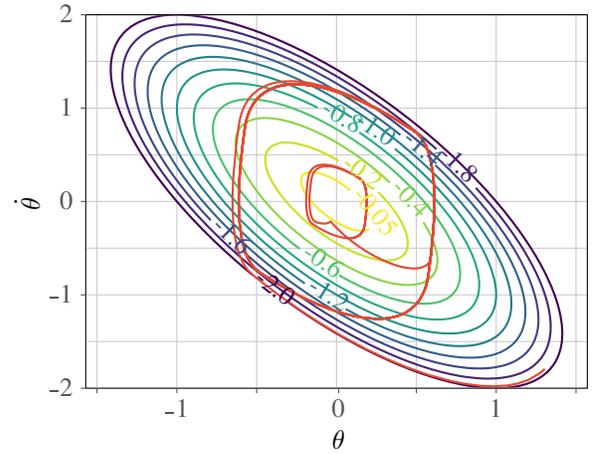
	\caption{Level sets of $ b $ and phase plot for $ t\in[0,20] $.}
	\label{fig:phase plot}
\end{figure}

\begin{figure}[t]
	\centering
	\def\svgwidth{0.85\columnwidth}
	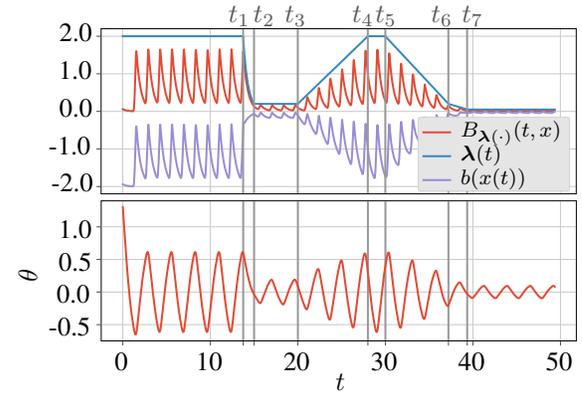
	\caption{Simulation results over time.}
	\label{fig:theta}
\end{figure}

In order to illustrate the application of our results, let us reconsider an example by Sontag \cite[Example 5.7.5]{Sontag1998} in a slightly modified form. In particular, let us consider the pendulum $\dot{x} = f(x,u)$, $ x = [x_{1}, \, x_{2}]^{T} \in \bbR^{2} $, detailed~as
\begin{align*}
	\dot{x}_{1} &= x_{2} \\
	\dot{x}_{2} &= -\frac{g}{l} \sin{(x_{1})} + d_{m}(x_{2}) + u
\end{align*} 
where $ x_{1} = \theta $ and $ x_{2} = \dot{\theta} $ are the excitation angle and velocity, respectively, $ u\in\calU\subseteq\bbR $ is the control input, $ d_{m}(x_{2}) = 5 l \, x_{2} $ is some destabilizing momentum, and we choose $ g=9.81 $ and $ l=1 $. 

As in \cite{Sontag1998}, we consider the CLF $ V(x) = 2 x_{1}^{2} + x_{2}^{2} + 2 x_{1} x_{2} $. It can be shown that for some $ u\in\calU = [-20,20] $ and all $ x $ with $ V(x) \leq 2 $ it holds 
\begin{align*}
	&-\frac{\partial V}{\partial x}(x) \, f(x,u) \geq \max\{2V(x), 2x_{2}^{2}\} \\
	&\quad \geq \gamma(V\!(x)) \!=\!
	\begin{cases}
		V(x) & \text{for } V(x) \!<\! 0.03 \\
		0.03 \!+\! 2 (V\!(x)\!-\!0.03) & \text{otherwise}
	\end{cases}
\end{align*}
where $ \gamma: \bbR_{\geq0}\rightarrow\bbR_{\geq0} $ is a class~$ \calK $ function. By Theorem~\ref{thm:clf and cbf}, we know that $ b(x) := -V(x) $ is a $ \Lambda $-shiftable CBF with $ \Lambda = 2 $; its level sets are depicted in Figure~\ref{fig:phase plot}. The extended class~$ \calK_{e} $ function $ \alpha $ corresponding to $ V $ is constructed from $ \gamma $ as in the proof of Theorem~\ref{thm:clf and cbf}. As $ \gamma $ is convex, we can choose the class~$ \calK $ function $ \alpha_{\lambda} $ as $ \alpha_{\lambda}(\xi) = -\alpha_{\lambda}(-\xi) = \gamma(\xi) $ for all $ \xi\in[0,\Lambda] $ (Sec.~\ref{subsec:lambda function construction guidelines}, step~1).

Now, we are ready for the construction of $ \bm{\lambda} $  (Sec.~\ref{subsec:lambda function construction guidelines}, step~2). We construct $ \bm{\lambda} $ as a continuous and piecewise continuously differentiable trajectory such that~\eqref{eq:lambda_dot condition} holds for all $ t\geq0 $; the resulting trajectory $ \bm{\lambda} $ is depicted in Figure~\ref{fig:theta}. In particular for $ t\in[t_{1},t_{2}] $, we construct $ \bm{\lambda} $ such that~\eqref{eq:lambda_dot condition} holds with equality, which requires to solve a differential equation. This yields the fastest possible decreasing $ \bm{\lambda} $ such that~\eqref{eq:lambda_dot condition} still holds. On the other intervals, we construct $ \bm{\lambda} $ as linear functions which allows to easily verify~\eqref{eq:lambda_dot condition} (e.g., for $ t\in[t_{5},t_{6}] $, $ \bm{\lambda} $ is such that $ \dot{\bm{\lambda}}(t) = \alpha_{\lambda}(\lambda(t_{6})) = \text{const.}$). Finally according to Theorem~\ref{thm:time-varying CBF without input constraints}, the uniformly time-varying CBF is given as $ B_{\bm{\lambda}(\cdot)}(t,x) := b(x) + \bm{\lambda}(t) $ (Sec.~\ref{subsec:lambda function construction guidelines}, step~3). 

Next, let us apply the feedback controller
\begin{align}
	\label{eq:invariance controller}
	\begin{split}
		&u(t,x) = \underset{u\in\calU}{\text{argmin}} \; |u| \\
		&\text{s.t. } \frac{\partial B_{\bm{\lambda}(\cdot)}}{\partial x}(t,x)\, f(x,u) + \frac{\partial \bm{\lambda}}{\partial t}(t) \geq -\beta(B_{\bm{\lambda}(\cdot)}(t,x))
	\end{split}
\end{align}
where $ \beta $ is constructed as in the proofs of Theorem~\ref{thm:time-varying CBF without input constraints} and Lemma~\ref{lemma:time-varying CBF without input constraints 1}. Note that $ u(t,x) $ is Lipschitz-continuous by~\cite[Theorem~3]{Ames2017}. Thus by Corollary~\ref{corollary:time varying set invariance with input constraints}, $ \calC_{\bm{\lambda}(\cdot)}(t) $ is forward invariant. The simulation results for $ x(0) = [1.3, \, -1.8] $ are depicted in Figure~\ref{fig:phase plot} (phase plot) and  Figure~\ref{fig:theta}. As $ B_{\bm{\lambda}(\cdot)} $ stays positive for all times, $ \calC_{\bm{\lambda}(\cdot)} $ is also rendered forward invariant in the simulation. Moreover, we note that the forward invariance of $ \calC_{\bm{\lambda}(\cdot)}(t) $ implies that the state constraints $ |\theta(t)|\leq0.45 $ and $ |\theta(t)|\leq0.225 $ are satisfied for $ t\in[t_{2},t_{3}] $ and $ t\geq t_{7} $, respectively, and that it holds $ |\theta(t)| \leq 1.42 $ for all times $ t\geq 0 $. More formally, in accordance with~\cite{Lindemann2019}, the above construction of the uniformly time-varying set $ \calC_{\bm{\lambda}(\cdot)} $ encodes the STL-specification
\begin{align*}
	\phi &:= \calG_{[t_{2},t_{3}]}(|\theta(t)| \leq 0.45) \wedge \calG_{[t_{7},\infty)}(|\theta(t)| \leq 0.225) \\
	&\quad \wedge \calG_{[0,\infty)}{(|\theta(t)| \leq 1.42)}
\end{align*}
where $ \calG $ denotes the always-operator and $ \wedge $ the logic AND (conjunction). In contrast to~\cite{Lindemann2019}, our method guarantees the existence of control inputs $ u $ satisfying $ u\in\calU $ even after the addition of the time-varying function $ \bm{\lambda} $ to the CBF $ b $. This is due to the following reasons: $ b $ is not only a CBF but it even satisfies the stronger properties of a $ \Lambda $-shiftable CBF, and $ \bm{\lambda} $ satisfies the differential inequality~\eqref{eq:lambda_dot condition}.

\section{Conclusion}
\label{sec:conclusion}

The subclass of CBFs, that we defined as $ \Lambda $-shiftable CBFs, has favorable properties for the satisfaction of uniformly time-varying constraints: starting from a time-invariant $ \Lambda $-shiftable CBF, various uniformly time-varying CBFs can be constructed. These can be used for the satisfaction of uniformly time-varying constraints as they for example occur in the literature on handling STL specifications. Advantageous about our approach is that a $ \Lambda $-shiftable CBF, once constructed, can be reused for various control objectives and it is not necessary to construct a new CBF for each change in the control objective.

\bibliographystyle{IEEEtran}
\bibliography{/Users/wiltz/CloudStation/JabBib/Research/000_MyLibrary}


\end{document}